\documentclass[showpacs,prb,twocolumn]{revtex4}
\RequirePackage{ifpdf}
\ifpdf
\pdfoutput=1
\else
\fi

\usepackage{amsmath}
\usepackage{amssymb}
\usepackage{color}
\usepackage{graphicx}
\usepackage{dcolumn}
\def\boldrm#1{{\bf #1}}

\def\q{\quad}

\begin{document}

\title {Electromagnetic wave transmission through a small hole in a perfect electric conductor of finite thickness}

\author{ A.~Yu.~Nikitin$^{1,2}$}
\email{alexeynik@rambler.ru}
\author{ D.~Zueco$^{1}$\footnote{Present address: Institut f\"{u}r Physik, Universit\"{a}t
Augsburg, Universit\"{a}tsstra{\ss}e 1, D-86135 Augsburg, Germany}}

\author{ F.~J.~Garc\'{i}a-Vidal$^3$}
\author{ L.~Mart\'{i}n-Moreno$^1$}
\email{lmm@unizar.es}
 \affiliation{$^1$ Instituto de Ciencia de Materiales de Arag\'{o}n and Departamento de F\'{i}sica de la Materia Condensada,
CSIC-Universidad de Zaragoza, E-50009, Zaragoza, Spain \\
$^2$ Theoretical Physics Department, A.Ya. Usikov Institute for Radiophysics and Electronics, Ukrainian Academy of
Sciences, 12 Acad. Proskura Str., 61085 Kharkov, Ukraine\\
$^3$ Departamento de F\'{i}sica Te\'{o}rica de la Materia Condensada, Universidad Aut\'{o}noma de Madrid, E-28049
Madrid, Spain}

\begin{abstract}
The non-resonant electromagnetic transmission of a normal-incident plane wave through a single hole in a perfect
conductor metal slab of finite width is studied. The cases of rectangular and circular holes are treated in detail. For
holes in the extreme subwavelength regime, in a film of finite thickness, the transmittance is shown to have the
Rayleigh dependency upon the wavelength and, in addition, is mainly suppressed due to attenuation of the fundamental
waveguide mode. In the limit of an infinitesimally thin screen Bethe's result is recovered for the circular hole. The
numerical computations are fitted, providing expressions for the transmission in a wide region of parameters. We
reformulate our results in terms of multipole expansion, interpreting the waveguide modes inside the hole as induced
multipole moments. This result provides the link between the modal expansion method and the one based on a multipole
expansion.
\end{abstract}

 \maketitle

\section{\label{sec:Intro}Introduction}

Electromagnetic (EM) wave transmission through apertures in perfect metal screens has been the subject of multiple
studies. For a long time, up to the middle of XX century, theoretical treatments of diffraction by either opaque or
metal bodies based on the Kirchhoff approach. This method consists in setting the fields on the body equal to their
incident values. Bethe was the first to consider the diffraction by a small circular aperture of radius $a$ in an
infinitesimally thin perfectly conducting screen, providing rigorous analysis of the Maxwell's equations with the exact
boundary conditions in 1944.\cite{Bethe} He found analytically the pre-factor appearing in the well-known Rayleigh
scattering dependency $\sim(a/\lambda)^4$ for the scattering cross-section of small objects in the long-wavelength
limit ($a\ll\lambda$). Later on, Bowkamp improved Bethe's result, providing additional terms in a series expansion of
the transmission over $a/\lambda$. Since then, this kind of diffraction problem has become a classical chapter in many
monographs on electromagnetism, e.g., Refs.~\onlinecite{Book_Jackson,Book_VanBladel}.

The interest in EM transmission through apertures has been renewed thanks to the discovery in 1998 of the enhanced
optical transmission through an array of small holes.\cite{Ebbesen98_Nature} A great deal of research has been devoted
to the transmission through periodical arrays of the holes. Nevertheless, the analysis of the diffraction by a single
hole in a film of finite thickness is still incomplete, as results found in the literature are only for fixed
geometrical parameters. For example, Refs.~\onlinecite{Roberts87,GAbajoOptExpr02} provide computed transmission spectra
through a circular hole in a film of a finite thickness in a perfect electric conductor (PEC) slab. The case of a
single circular hole in a real metal has been considered in
Refs.~\onlinecite{WannemacherOptCom01,PopovApplOpt05,SchatzOptExpr05}. However, it is problematic to extrapolate these
results to other parameters and other hole shapes. An attempt to represent the solution in an analytical form was
undertaken in Ref.~\onlinecite{ShalaevSarych05}. These authors derived the normalized cross section for a circular
aperture (of radius up to half of the wavelength in PEC) assuming that the magnetic current is uniform within the
aperture. Nevertheless, this model shows poor quantitative agreement with both the Bethe-Bouwkamp results and the
strict numerical calculations of Roberts.\cite{Roberts87} Moreover, it is limited to a screen with zero thickness.

In this paper we study the optical transmission through single holes in PEC. We present new analytical results, valid
for a wide range of geometrical parameters, and provide a link between the modal expansion method and the one based on
multipoles. The paper is organized as follows: In Section II we describe the modal expansion technique used for
studying the transmission through the hole in a perfect electric conductor film of arbitrary thickness. In Section III
we provide analytical expressions for the transmission through holes of both rectangular and circular shapes in extreme
subwavelength limit. We test our approach by applying it to the most unfavorable conditions for the method used (when
the film thickness is zero), and obtain an excellent agrement with known results. In Section IV the square hole of
moderate size is treated, when the wavelength is still larger than the hole cutoff (therefore, the resonances found
close to cutoff\cite{hole_shape_PRL04,Ebbesen_singHole_OptCom04,hole_shape_PRL07} will not be discussed there, as have
already been addressed before\cite{rect_hole_PRL05,resonant_hole_PRB06}). Finally, in Section V, we make the link
between the induced multipoles and the waveguide modes inside the hole. We discuss the importance of the fundamental
waveguide mode both for the formation of the induced dipole moments on both faces of the hole and for the coupling
between these moments. In Appendix we explain the simplifications of Green's tensor for the small hole limit.

\section{\label{sec:math}Theoretical background}

\begin{figure}[h!]
  \includegraphics[width=7cm]{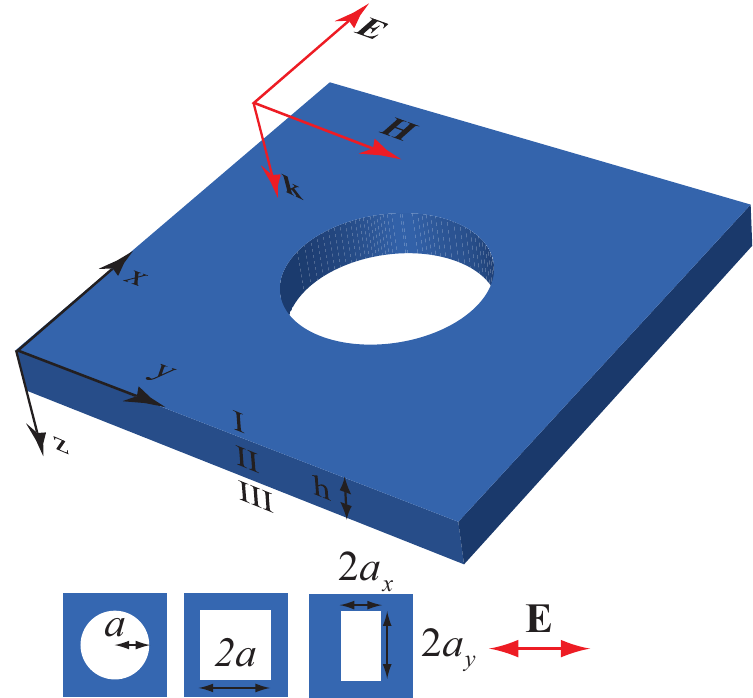}\\
  \caption{(Color online) Geometry of the problem.}\label{geom}
\end{figure}


In this section we briefly outline the modal expansion formalism for the EM field.\cite{JorgePRL2004,rect_hole_PRL05}
Consider an EM wave incident onto a PEC film of finite thickness $h$ containing a hole, see Fig.\ref{geom}. We first
assume that both the bounding media and the medium inside the aperture is vacuum, but later on we generalize the
results to arbitrary values of the optical indices of the semi-infinite media.

For the representation of the EM field in-plane components we use Dirac's notations as follows. The projection of a
given Dirac's vector onto the position vector $|\mathbf{r}_t\rangle$, with $\mathbf{r}_t=(x,y)$, simply yields the
value of Dirac's vector in the position $\mathbf{r}_t$ coordinate dependent field in the coordinate space, for example:
\begin{eqnarray}\label{tb3}
\langle \mathbf{r}_t|\mathbf{E}\rangle = \mathbf{E}_t(x,y).
\end{eqnarray}
The projection of a vector onto another one is given by the scalar product
\begin{eqnarray}\label{tb4}
\langle \alpha|\kappa\rangle = \int d\mathbf{r}_t \langle \alpha|\mathbf{r}_t\rangle \langle
\mathbf{r}_t|\kappa\rangle^\ast,
\end{eqnarray}
where ``$\ast$'' means complex conjunction.

Let us write in Dirac's notations the tangential components of the fields in the lower and upper half-spaces, expanding
them over the continuum of plane waves (see the geometry in Fig.~\ref{geom}):
\begin{eqnarray}\label{tb1}
&&|\mathbf{E}_I(z)\rangle = e^{ik_{zi}z}|\kappa_i\rangle  + \sum\limits_\kappa r_\kappa e^{-ik_zz}|\kappa\rangle, \nonumber\\
&&|\mathbf{E}_{III}(z)\rangle = \sum\limits_\kappa t_\kappa e^{ik_z(z-h)}|\kappa\rangle.
\end{eqnarray}
Here $\kappa = (\mathbf{k},\sigma)$ represents both the in-plane wavevector component $\mathbf{k} = (k_x,k_y)$ and the
polarization, $\sigma=p$ or $\sigma=s$. The coordinate representation of the modes in the vacuum half-spaces reads
\begin{eqnarray}\label{tb1.1}
\langle \mathbf{r}_t|\mathbf{k},s\rangle =
\begin{pmatrix}
-k_y\\
k_x
\end{pmatrix}\frac{e^{i\mathbf{k}\mathbf{r}_t}}{k}, \q
\langle \mathbf{r}_t|\mathbf{k},p\rangle =
\begin{pmatrix}
k_x\\
k_y
\end{pmatrix}\frac{e^{i\mathbf{k}\mathbf{r}_t}}{k}.
\end{eqnarray}
The plane wave propagation constant is $k_z = \sqrt{g^2-k^2}$ with $g = 2\pi/\lambda$. The summation operator in
Eq.~\eqref{tb1} includes both the integration over the $k$ continuum spectrum and the summation over the polarizations:
$\sum_\kappa=(1/2\pi)^2\sum_\sigma\int d\mathbf{k}$. Inside the hole, we expand the tangential components of the field
over the modes $|\alpha\rangle$ of the infinite waveguide
\begin{eqnarray}\label{tb2}
&|\mathbf{E}(z)\rangle = \sum\limits_{\alpha}\left(A_\alpha e^{iq_{z\alpha}z}+B_\alpha
e^{-iq_{z\alpha}z}\right)|\alpha\rangle,
\end{eqnarray}
where $A_\alpha$ and $B_\alpha$ are the amplitudes of the waveguide modes propagating (or decaying) forwardly and
backwardly with respect to $z$-axis direction; $q_{z\alpha}$ represents the propagation constant of a waveguide mode
with the label $\alpha$. This label contains both the polarization of the waveguide mode and a ``spatial'' index
related to the number of nodes of the field inside the hole.

By matching the EM fields at the interfaces, and using the orthogonality of the modes, we arrive at a set of linear
equations for the expansion coefficients
\begin{eqnarray}\label{tb5}
&&(G_{\alpha\alpha}-\Sigma_\alpha)E_\alpha+\sum\limits_{\alpha\neq\beta}G_{\alpha\beta}E_\beta-
G_{\alpha}^VE'_\alpha = I_\alpha, \nonumber\\
&&(G_{\gamma\gamma}-\Sigma_\gamma)E'_\gamma+\sum\limits_{\nu\neq\gamma}G_{\gamma\nu}E'_\nu- G_{\gamma}^VE_\gamma = 0.
\end{eqnarray}
The coefficients $E_\alpha$ and $E'_\alpha$ are
\begin{eqnarray}\label{tb6}
&&E_\alpha=A_\alpha+B_\alpha, \nonumber\\
&&E'_\alpha=-\left(A_\alpha e^{iq_{z\alpha}h}+B_\alpha e^{-iq_{z\alpha}h}\right),
\end{eqnarray}
so that the system of Eqs.~\eqref{tb5} connects the electric field modal amplitudes on the incoming interface, $z=0$,
and on the outgoing one, $z=h$. The term $G^V_\alpha$ describes the coupling between the input and output sides of the
holes, and $\Sigma_\alpha$ arises from the reflection of the waveguide mode at the openings:
\begin{eqnarray}\label{tb7}
G^V_\alpha = \frac{2iY_\alpha e^{iq_{z\alpha}h}}{e^{2iq_{z\alpha}h}-1}, \q \Sigma_\alpha =
iY_\alpha\frac{e^{2iq_{z\alpha}h}+1}{e^{2iq_{z\alpha}h}-1},
\end{eqnarray}
where $Y_\alpha  = q_{z\alpha}/g$ is the admittance for the TE waveguide mode, and $Y_\alpha  = g/q_{z\alpha}$ is that
for the TM one. The coupling matrix elements of the system of Eqs.~\eqref{tb5} are related to the in-plane components
of the EM Green function dyadic $\hat{G}$. The latter is associated to a homogeneous medium in three dimensions and is
represented in the waveguide mode space:
\begin{eqnarray}\label{tb8}
G_{\alpha\beta} = \langle \alpha|\hat{G}|\beta\rangle = i\sum\limits_\kappa Y_\kappa \langle
\alpha|\kappa\rangle\langle\kappa|\beta\rangle.
\end{eqnarray}
Here $Y_\kappa$ is the admittance of the mode in free space: $Y_{\mathbf{k},s}=k_z/g$ and $Y_{\mathbf{k},p}=g/k_z$. The
right-hand side (r.h.s.) term $I_\alpha$ takes into account the overlap between the incident plane wave and the
waveguide mode $|\alpha\rangle$ inside the hole. Considering a normalization for the incident wave such that the energy
flux through the hole is unity, $\mathrm{Re}(\int_{hole} d\mathbf{S}\, \mathbf{E}_i\times\mathbf{H}_i^\ast) = 1$, we
obtain
\begin{eqnarray}\label{tb9}
I_{\alpha} = 2i \sqrt{Y_{\kappa_i}} \langle\kappa_i|\alpha\rangle,
\end{eqnarray}
where $Y_{\kappa_i}$ is the admittance of the incident plane wave in free space.

Once the solution of the system of Eqs.~\eqref{tb5} is found and the modal amplitudes are known, the normalized-to-area
transmission coefficient can be written as
\begin{eqnarray}\label{tb10}
T = \sum\limits_{\alpha,\beta}\mathrm{Im}(G_{\alpha\beta})E'_\alpha E'^\ast_\beta.
\end{eqnarray}

Until now we have not mentioned any restrictions on the hole shape, which in our formalism only influences the
structure of the waveguide modes $|\alpha\rangle$. In this paper, however, we restrict ourselves to the consideration
of both circular and rectangular holes, where the waveguide modes are known analytically.

For thick films the solution of Eqs.~\eqref{tb5} converges quickly. The reason is that in the subwavelength limit the
amplitudes of the waveguide modes decay inside the hole. As the decay is characterized by the propagation constants
$q_\alpha$, the higher the waveguide mode index, the weaker its influence on the transmission. Therefore, only a few
waveguide modes (with the smallest decrements) contribute the transmission.

In contrast, when the thickness $h$ of the PEC tends to zero, the solution of the system of Eqs.~\eqref{tb5} involves
many waveguide modes (hundreds or even thousands) to provide the precise result. However, we shall show below that a
very accurate computation of the transmission can be performed with some tens of waveguide modes. We shall find the
asymptotic value of the transmission by a fitting of the convergent result.

To conclude this section we shall show how to apply the above equations when the bounding dielectric media have
permittivities $\epsilon_I$, $\epsilon_{III}$.  In this case, for a medium with the dielectric constant $\epsilon$, the
propagation constant of the mode and the admittances become $k_z = \sqrt{\epsilon g^2-k^2}$, $Y_{\mathbf{k},s}=k_z/g$
and $Y_{\mathbf{k},p}=\epsilon g/k_z$. Then the wavelength-dependent tensor $G_{\alpha\beta}=G_{\alpha\beta}(\lambda)$
describing the interaction of the waveguide modes inside the cavity through the EM continuum in vacuum becomes a
function of $\epsilon$, $G_{\alpha\beta}(\lambda;\epsilon)$. In the upper equation of \eqref{tb5} $G_{\alpha\beta}$
changes to $G_{\alpha\beta}(\lambda;\epsilon_I)$, whereas in the lower to $G_{\alpha\beta}(\lambda;\epsilon_{III})$.
Working out the expression given by Eq.~\eqref{tb8}, we find that the relation between the tensors is
\begin{eqnarray}\label{tb11}
G_{\alpha\beta}(\lambda;\epsilon) = \sqrt{\epsilon}G_{\alpha\beta}(\lambda/\sqrt{\epsilon};\epsilon=1).
\end{eqnarray}

\section{\label{sec:small}Small hole limit}

In this section we show how to simplify computation of the transmission when the linear size of the hole,
$a\sim\sqrt{S}$ is small compared to $\lambda$. More precisely, we consider the limit
\begin{eqnarray}\label{sh1}
\varepsilon\equiv ga\ll1.
\end{eqnarray}
In this limit an accurate numerical computation of the tensor $G_{\alpha\beta}$ becomes problematic. This is related to
the orders of magnitude difference between imaginary and real parts of $G_{\alpha\beta}$. We have found that in the
low-order in parameter $\varepsilon$ the non-vanishing elements of the tensor depend upon $\varepsilon$ as  (see
Appendix \ref{GT})
\begin{eqnarray}\label{sh2}
\mathrm{Im}(G_{\alpha\beta})\sim \varepsilon^2, \q \mathrm{Re}(G_{\alpha\beta})\sim 1/\varepsilon.
\end{eqnarray}
Both real and imaginary parts of the tensor are important in spite of their substantial difference:
$\mathrm{Re}(G_{\alpha\beta})$ contribute into the amplitude of the waveguide modes (see below), while
$\mathrm{Im}(G_{\alpha\beta})$ takes into account the radiation into free-space [see Eq.~\eqref{tb10}]. For
$\varepsilon\ll1$ the imaginary part of the tensor allows analytical computation, and the real part can be considerably
simplified (see details in Appendix \ref{GT}).

In this paper we restrict ourselves to normal-incident wave transmission. In this case the impinging wave can only
couple to certain waveguide modes of $TE$ type. The analysis of the $G_{\alpha\beta}$ elements shows that the
contribution to the transmission from $TM$ waveguide modes is always negligible. This results from both a weak coupling
between the waveguide modes of different polarizations inside the small hole, and a weak coupling of $TM$ waveguide
modes to the EM continuum of vacuum half-spaces.

For the circular hole the incident plane wave couples directly with only the ``horizontal'' $TE_{1n}$ waveguide modes
with integer $n$ (where the first index indicates the number of semi-periods of the field placed along the polar
angle). Coupled between themselves, these waveguide modes are the only ones contributing into the transmission.

For a rectangular hole, when the electric field is directed as shown in Fig.~\ref{geom}, the illuminated waveguide
modes are $TE_{0n}$, with odd $n$ (the first and the second indices define the number of semi-periods of the field
placed along $x$ and $y$ directions respectively). Only this set of waveguide modes appears in the summation for the
transmission according to Eq.~\eqref{tb10} [all other elements of $\mathrm{Im}(G_{\alpha\beta})$ are negligible].
However, $TE_{0n}$ modes couple to $TE_{mn}$ ones with even $m$ and odd $n$ through the system of Eqs.~\eqref{tb5} and
must be taken into account.

Due to property expressed in Eq.~\eqref{sh2}, in the extreme subwavelength limit the transmission coefficient scales as
\begin{eqnarray}\label{sh3}
T = \varepsilon^4\psi(h).
\end{eqnarray}
The thickness- and shape-dependent function $\psi(h)$ is given by the solution of the system of Eqs.~\eqref{tb5} with
appropriately normalized coefficients
\begin{eqnarray}\label{sh4}
\psi(h) = \sum\limits_{\alpha,\beta}\tilde{G}_{\alpha\beta}\tilde{E}_\alpha(h) \tilde{E}^\ast_\beta(h),
\end{eqnarray}
where
\begin{eqnarray}\label{sh5}
\tilde{G}_{\alpha\beta} = \frac{\mathrm{Im}(G_{\alpha\beta})}{\varepsilon^2} \q \mathrm{and} \q \tilde{E}_\alpha =
\frac{E'_\alpha}{\varepsilon}.
\end{eqnarray}
The amplitudes $\tilde{E}_\alpha$ satisfy the system of Eqs.~\eqref{tb5}, where the imaginary part of the Green tensor
is neglected and its real part must be normalized with the small parameter $\mathrm{Re}(G_{\alpha\beta})\rightarrow
\varepsilon\mathrm{Re}(G_{\alpha\beta})$. The coefficients of Eq.~\eqref{tb7} are replaced by $G^V_\alpha \rightarrow
\varepsilon G^V_\alpha $, $\Sigma_\alpha \rightarrow \varepsilon \Sigma_\alpha $.

\subsection{\label{subsec:scren}Perfect electric conductor screen}

In this subsection we compare the solution based on our formalism with some known results on the transmission through
apertures in an infinitesimally thin PEC screen. By reaching an excellent agreement with these results, we justify the
applicability of the modal expansion even in the most unfavorable conditions for it.

In the limit $h\rightarrow 0$, special care is needed when solving system of Eqs.~\eqref{tb5} due to the divergency of
the coefficients $G^V_\alpha$ and $\Sigma_\alpha$. In order to remove this divergency, we expand the amplitudes of the
waveguide modes over $gh$
\begin{eqnarray}\label{shs1}
& E_\alpha = E_\alpha^{(0)} + (gh)E_\alpha^{(1)} + ...,\nonumber\\
& E'_\alpha = E_\alpha'^{(0)} + (gh)E_\alpha'^{(1)} + ...
\end{eqnarray}
Equating terms proportional to $(gh)^{-1}$, we obtain
\begin{eqnarray}\label{shs2}
E_\alpha^{(0)} =-E_\alpha'^{(0)}.
\end{eqnarray}
Then, keeping terms of zero-order in $gh$, we have from Eqs.~\eqref{tb5}
\begin{eqnarray}\label{shs2.1}
&&\sum\limits_{\beta}G_{\alpha\beta}E^{(0)}_\beta-
Y_{\alpha}(E^{(1)}_\alpha+E'^{(1)}_\alpha) = I_\alpha, \nonumber\\
&&\sum\limits_{\beta}G_{\alpha\beta}E^{(0)}_\beta + Y_{\alpha}(E^{(1)}_\alpha+E'^{(1)}_\alpha) = 0.
\end{eqnarray}
Adding the two equations in \eqref{shs2.1} and neglecting the imaginary part of the tensor $G_{\alpha\beta}$, due to
Eq.~\eqref{sh2}, we arrive at the final system of equations
\begin{eqnarray}\label{shs3}
2\sum\limits_{\beta}\mathrm{Re}(G_{\alpha\beta})E_\beta^{(0)} =I_\alpha.
\end{eqnarray}
Now, when the film has been converted into a screen with infinitesimal thickness, the amplitudes of the waveguide modes
at the incoming and outgoing faces of the film are equivalent. Therefore, the terms $\Sigma_\alpha$ and $G^V_\alpha$
responsible for the reflection and coupling of the waveguide modes inside the cavity are not present in
Eq.~\eqref{shs3}, and the waveguide mode amplitudes are coupled by the doubled elements of
$\mathrm{Re}(G_{\alpha\beta})$.

Using the normalization defined by Eqs.~\eqref{sh4}, \eqref{sh5}, the transmission coefficient can be written in the
form \eqref{sh3}, where the function $\psi(h)$ becomes a constant defined by the shape of the hole.
\begin{eqnarray}\label{shs4}
T = \varepsilon^4 C.
\end{eqnarray}

The simplest solution of the system of Eqs.~\eqref{shs3} is obtained by retaining only the fundamental waveguide mode.
For the circular hole, the Green tensor element corresponding to the fundamental waveguide mode is $G_{TE_{11}TE_{11}}
= 1.1951/\varepsilon+0.2789i\varepsilon^2$.  For the square hole $G_{TE_{01}TE_{01}} =
0.9577/\varepsilon+0.344i\varepsilon^2$. Within this single-mode approximation, taking the r.h.s. from Appendix A, the
transmission pre-factors of Eq.~\eqref{shs4} are found immediately: for the circular hole $C_\circ=0.1634$ and for the
square one $C_\square=0.3041$. For a circular hole, this minimal model provides the transmission of order of $30$\%
with respect to the exact Bethe's result $C_\circ = 64/(27\pi^2)\simeq0.2402$. Therefore, more waveguide modes must be
taken into account in order to obtain the correct result.

\begin{figure}[h!]
  \includegraphics[width=7cm]{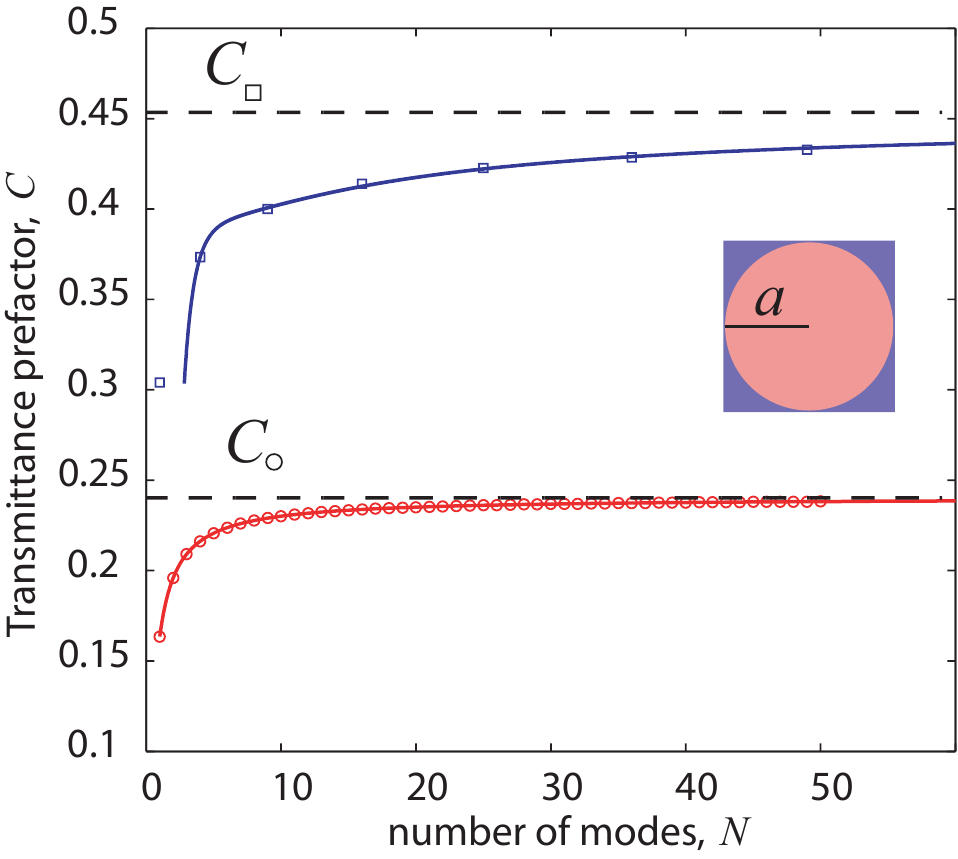}\\
  \caption{(Color online) Normalized by $(ga)^4$ transmittance for the circular and
square apertures of the size $a$ in the PEC screen. The asymptotic values are shown by the dashed
lines.}\label{converg}
\end{figure}

The convergency of the constant for the circle, $C_\circ$, and for the square, $C_\square$, with respect to the number
of waveguide modes are shown in Fig.~\ref{converg}. After having computed the value of this constant for several tens
of waveguide modes, we fit it by a polynomial
\begin{eqnarray}\label{shs5}
C = \sum_{m=0}^{m_{max}}\frac{a_m}{N^m},
\end{eqnarray}
where $N$ is the number of the waveguide modes.

The value $a_0$ gives us the constant $C$. In the calculations shown in Fig.~\ref{converg}, the fitting has been done
by a forth-order polynomial ($m_{max} = 4$) using 50 waveguide modes. For the circular hole we obtain $C_\circ=a_0
=0.2403$ with an error of only $0.05$\% with respect to the exact Bethe's value. For the square hole we obtain
$C_\square\simeq 0.4565$. Notice that the normalized transmittance through a square hole with the side $2a$ is about
two times larger than that of the round hole with the radius $a$.

\begin{figure}[h!]
  \includegraphics[width=7cm]{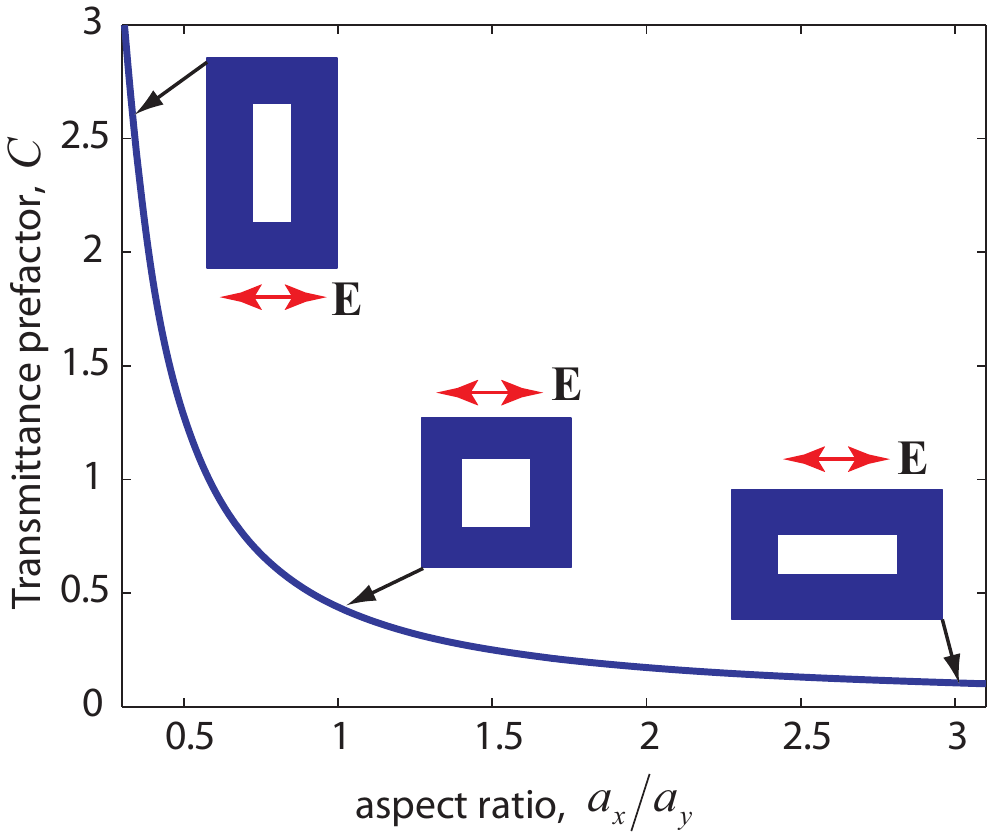}\\
  \caption{(Color online) Transmittance for the rectangular aperture in the PEC screen,
normalized by $(ga_x)^2(ga_y)^2$. The shape of the rectangle for $a_x/a_y$ equal to $1/3$, $1$ and $3$ is
presented.}\label{rect}
\end{figure}

In the case of rectangular holes, it is useful to write the transmittance as $T =
\varepsilon_x^2\varepsilon_y^2C(\tau)$, where $\varepsilon_{x} = a_{x}g$, $\varepsilon_{y} = a_{y}g$ and $\tau$ is the
aspect ratio
\begin{eqnarray}\label{shs4.1}
\tau = a_x/a_y.
\end{eqnarray}
In this representation $C(\tau)$ reflects the dependency of the transmittance upon the aspect ratio for a constant area
of the hole. As seen from Fig.~\ref{rect}, this dependency is a fast function of $\tau$.
This is due to a strong dependency of the polarizability of the hole upon the aspect ratio.
\cite{Polariz_exper_IEEE,Polariz_teor_IEEE} From the point of view of the modal expansion formalism, the cutoff
wavelength of the fundamental waveguide mode is $\lambda_c = 4a_y$, when the field is parallel to $x$-axis. Therefore,
the larger the $a_y$, the closer the hole to the resonant regime.\cite{rect_hole_PRL05}

We have fitted the dependency shown in Fig.~\ref{rect} in the region $\tau \in [1/3,3]$ by the following function
\begin{eqnarray}\label{shs5}
C(\tau) = 0.0132 + 0.2127/\tau+ 0.2174/\tau^2.
\end{eqnarray}
This fitted function provides an excellent approximation to the transmittance: in Fig.~\ref{rect} the curve given by
Eq.~\eqref{shs5} is indistinguishable from that obtained from the numeric calculations. In the interval $\tau<1$ the
dependency $C(\tau)$ can be extracted from Ref.~\onlinecite{Polariz_teor_IEEE}, and we have checked that it coincides
with Eq.~\eqref{shs5}.

Let us now turn to the dependency of the transmittance upon dielectric permittivities of the bounding media. Addressing
to the general property of $G_{\alpha\beta}$ in Eq.~\eqref{tb11} and to the scaling given by Eq.~\eqref{sh2}, where
$\varepsilon\sim1/\lambda$, we see that the imaginary part of the tensor is a function of $\epsilon_l$, namely
$\mathrm{Im}(G^l_{\alpha\beta})\sim\epsilon_l^{3/2}$, where $l=I,III$. In contrast, $\mathrm{Re}(G^l_{\alpha\beta})$ is
not dependent upon $\epsilon_l$, so that the amplitudes of the waveguide modes depend upon $\epsilon_I$ only through
the r.h.s., $E_\alpha,E'_\alpha\sim \sqrt{Y_{\kappa_i}}=\epsilon_I^{1/4}$. Then it follows directly from
Eq.~\eqref{tb10} that the transmittance for arbitrary substrate and superstrate is related to the one when the system
is in vacuum as
\begin{eqnarray}\label{shs6}
T(\lambda,\epsilon_I,\epsilon_{III})= \sqrt{\epsilon_I\epsilon_{III}^3}T(\lambda,\epsilon_I=1,\epsilon_{III}=1)
\end{eqnarray}
for $\varepsilon\sqrt{\epsilon_{I}}\ll1$, $\varepsilon\sqrt{\epsilon_{III}}\ll1$. Note that the transmission is not
symmetric with respect to the region of incidence: $T(\lambda,\epsilon_I,\epsilon_{III})\neq
T(\lambda,\epsilon_{III},\epsilon_I)$. This may seem to be paradoxical, as the transmission coefficient of the incident
plane wave into the plane wave with the same in-plane wavevector is symmetric due to the symmetry of the scattering
matrix. The total transmittance $T(\lambda)$, however, takes into account the transmission of a plane wave into a
\emph{continuum} of states. As the density of final states depends on the dielectric constant in the transmission
region, so does $T(\lambda)$. The integration over the scattering amplitudes yields a factor proportional to the
wavevector squared modulus in the transmitted medium, $T\sim (\sqrt{\epsilon_{III}}g)^2\sim\epsilon_{III}$, and breaks
the $I$-$III$ symmetry in Eq.~\eqref{shs6}.

\subsection{\label{subsec:thick}Thick and medium films}

The EM fields inside a subwavelength hole decay exponentially with both the film thickness and the propagation
constants of the waveguide modes. In the limit of a very thick film, $e^{-2|q_{z\alpha}|h}\ll1$, the coefficients of
Eq.~\eqref{tb7} are simplified and the solution of the system of Eqs.~\eqref{tb5} can be cast in the matrix form
\begin{eqnarray}\label{th2}
\hat{E} \simeq \hat{D}^{-1}\hat{I}, \q \hat{E}'\simeq \hat{D}^{-1}\hat{d}\hat{D}^{-1}\hat{I},
\end{eqnarray}
where
\begin{eqnarray}\label{th3}
\hat{D} = \|G_{\alpha\beta}+i\delta_{\alpha,\beta}Y_\alpha \|, \q \hat{d} = -2i\|\delta_{\alpha,\beta}Y_\alpha
e^{iq_{z\alpha}h} \|.
\end{eqnarray}
Thus, for very thick films the waveguide mode amplitudes on the input side do not depend upon $h$, and are of the same
order as the incident field. Conversely, the field amplitudes on the output side are exponentially decreased.

For very thick films only the fundamental $TE$ waveguide mode is expected to contribute into the coupling between both
sides of the hole. Therefore, we can retain in the diagonal matrix $\hat{d}$ only the element corresponding to the
fundamental waveguide mode. This means that for a thick film the transmittance decays as $T\sim e^{-2|q_{z0}|h}$, where
$q_{z0}$ is the propagating constant of the fundamental waveguide mode. For a rectangular hole with the sides $2a_x$
and $2a_y$, $q_{z0} =\sqrt{g^2-(\pi/2a_y)^2}\simeq i\pi/2a_y $, and for a circular hole of the radius $a$, $q_{z0}
=\sqrt{g^2-(u_{1}/a)^2}\simeq iu_{1}/a$ (see the definition of $u_{m}$ in Appendix~\ref{AIch}). Then for arbitrary film
thickness it is useful to rewrite $T$ given by Eq.~\eqref{sh3} in the following form
\begin{eqnarray}\label{th1}
T = \varepsilon^4e^{-2|q_{z0}|h}C(h),
\end{eqnarray}

\begin{figure}[h!]
  \includegraphics[width=7.5cm]{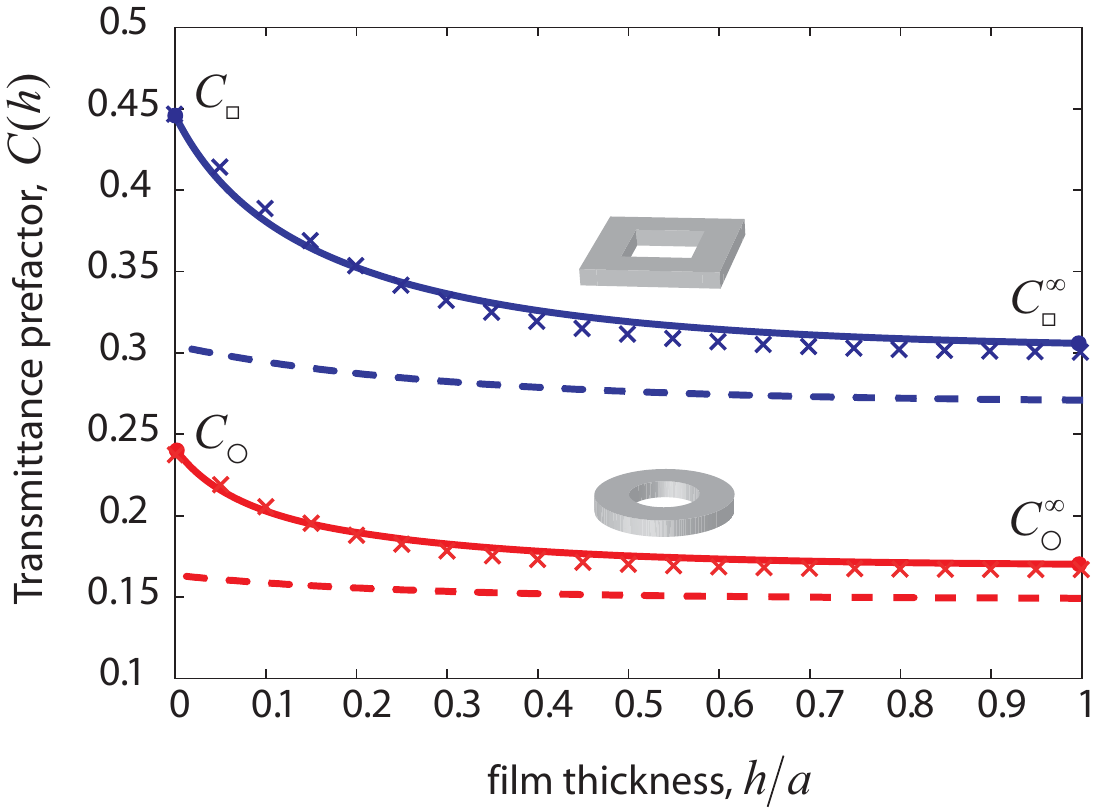}\\
  \caption{(Color online) The normalized transmission through small circular and
square holes as function of the film thickness. Calculations using only the fundamental waveguide mode are represented
by dashed lines. The values for the fitted expression \eqref{th4} are marked by ``$\times$''.}\label{h-depend}
\end{figure}

The computations of $C(h)$ in the interval $h/a\in[0,1]$ for both the square and circular holes are shown in
Fig.~\ref{h-depend}.
In both cases we phenomenologically adjust this dependency by the function
\begin{eqnarray}\label{th4}
C(h) = C^\infty+\left(C-C^\infty\right)e^{- \delta h/a},
\end{eqnarray}
containing fitting parameters $C$, $C^\infty$ and $\delta$. In the limit of a screen, $h=0$, Eq.~\eqref{th1} transforms
into Eq.~\eqref{shs4}, while for the infinite film thickness $f(h)$ becomes a constant $C^\infty=C(\infty)$. The
constants $C$ have been given in Subsection~\ref{subsec:scren}. The values of the constants for the infinite film
thickness have been found to be $C^\infty_\circ\simeq 0.1694$ and $C^\infty_\square\simeq 0.3027$. The values for
parameters $\delta$ are $\delta_\circ\simeq 6$ and $\delta_\square\simeq 5$.

For the rectangle the constants $C$ and $C^\infty$ in Eq.~\eqref{th4} are functions of the aspect ratio. $C = C(\tau)$
is given by Eq.~\eqref{shs5}, and $C^\infty=C^\infty(\tau)$ reads
\begin{eqnarray}\label{th5}
C^\infty(\tau) = 0.2298/\tau+ 0.08262/\tau^2.
\end{eqnarray}
Thus, the formula \eqref{th1} is applicable for the rectangle as well, if we replace
$\varepsilon\rightarrow\sqrt{\varepsilon_x\varepsilon_y}$ in Eq.~\eqref{th1} and $h/a\rightarrow h/a_y$ in
Eq.~\eqref{th4}  (we find that $\delta$ for the rectangular hole coincides with that for the square hole
$\delta_\square$).

In order to make our results more accessible, we summarize in Table~\ref{tab:table1} the constants appearing in
Eqs.~\eqref{th1} and \eqref{th4}.

As we see from the dependencies shown in Fig.~\ref{h-depend}, the difference between the full calculation and
single-mode approximation deceases when $h$ increases. For the zero thickness screen it is of order of $30 \%$, and for
the limit $h\rightarrow\infty$ it is of order of $10 \%$. However, if we extract the amplitude of the fundamental
waveguide mode from the full (many-mode) solution, and use only this waveguide mode to compute the transmission
according to Eq.~\eqref{tb10}, the difference between this calculation and the exact value is reduced.

\begin{table}[h!]
\caption{\label{tab:table1}Resume of the parameters appearing in the analytical representation of the transmission, see
Eqs.~\eqref{th1}, \eqref{th4}. }
\begin{ruledtabular}
\begin{tabular}{cccc}
 &Circular hole & Square hole & Rectangular hole\\
  &(radius=$a$) & (side=$2a$) & (sides=$2a_x,2a_y$, $\tau=\frac{a_x}{a_y}$)\\
\hline
\\
$C$ & $\frac{64}{27\pi^2}$ (Bethe\cite{Bethe}) & 0.4565 & $0.0132 + \frac{0.2127}{\tau}+ \frac{0.2174}{\tau^2}$\\
\\
$C^\infty$ & 0.1694 & 0.3027 & $\frac{0.2298}{\tau}+ \frac{0.08262}{\tau^2}$\\
\\
$|q_{z0}|$ & $\sqrt{(\frac{u_{1}}{a})^2-g^2}$ & $\sqrt{(\frac{\pi}{2a})^2-g^2}$ & $\sqrt{(\frac{\pi}{2a_y})^2-g^2}$\\
$\delta$ & 6 & 5 & 5\\
\end{tabular}
\end{ruledtabular}
\end{table}

To summarize this section, in the extreme subwavelength regime, many waveguide modes are necessary to provide the
precise value for the transmittance. On the other hand, the fundamental waveguide mode plays a crucial role in the
process, especially in the coupling between the fields on top and bottom faces of the hole. As we will show, this
reflects the fact that the fundamental waveguide mode possesses the largest induced dipole moment (see
Section~\ref{sec:multipole} below).

\section{\label{sec:largeholes}Holes of a moderate size}

The sizes of apertures used in the experimental samples are often not in the extreme subwavelength limit considered in
the previous section. For example, in original experiments on enhanced optical transmission\cite{Ebbesen98_Nature} the
sizes of the holes were of order of $200-300$ nm, so the parameter $\varepsilon$ was $\varepsilon\gtrsim1$ in the
visible range of the spectra.

In this section we study holes of moderate sizes: still in the subwavelength limit, but with the condition \eqref{sh1}
not fulfilled. However, we still consider wavelengths larger than the resonant wavelength of the hole. For the resonant
transmission through a single hole at wavelengthes close to the cutoff, we refer the reader to
Ref.~\onlinecite{rect_hole_PRL05}.

When the size of the hole increases, the transmittance value can be refined by retaining the next terms in the
expansion over $\varepsilon$. For a circular aperture in the PEC screen of zero thickness a few first terms  were
computed by Bouwkamp.\cite{Bouwkamp54} However, such a series has radius of convergence $R_\varepsilon\sim1$, and
therefore is applicable in a narrow region of $\varepsilon$.

For arbitrary film thickness and size of the hole, the transmission can be accurately computed numerically using the
modal expansion. But from the practical point of view it is useful to have an analytical formula containing the
parameters of the hole. We have numerically computed the transmission through a square hole retaining many waveguide
modes in the system of Eqs.~\eqref{tb5} in a wide range of the size $2a$, and the thickness $h$. Then we have
approximated the calculations generalizing dependency \eqref{th1} by adding a quadratic in $\varepsilon$ term to the
function $C(h)$
\begin{eqnarray}\label{ms1}
T = \varepsilon^4e^{-2|q_{z0}|h}[C(h)+\varepsilon^2C_2(h)],
\end{eqnarray}
where $C_2(h)$ has the same form that $C(h)$ in Eq.~\eqref{th4}
\begin{eqnarray}\label{ms2}
C_2(h)= C_2^\infty+\left(C_2-C_2^\infty\right)e^{-\delta h/a}.
\end{eqnarray}

\begin{figure}[h!]
  \includegraphics[width=7.5cm]{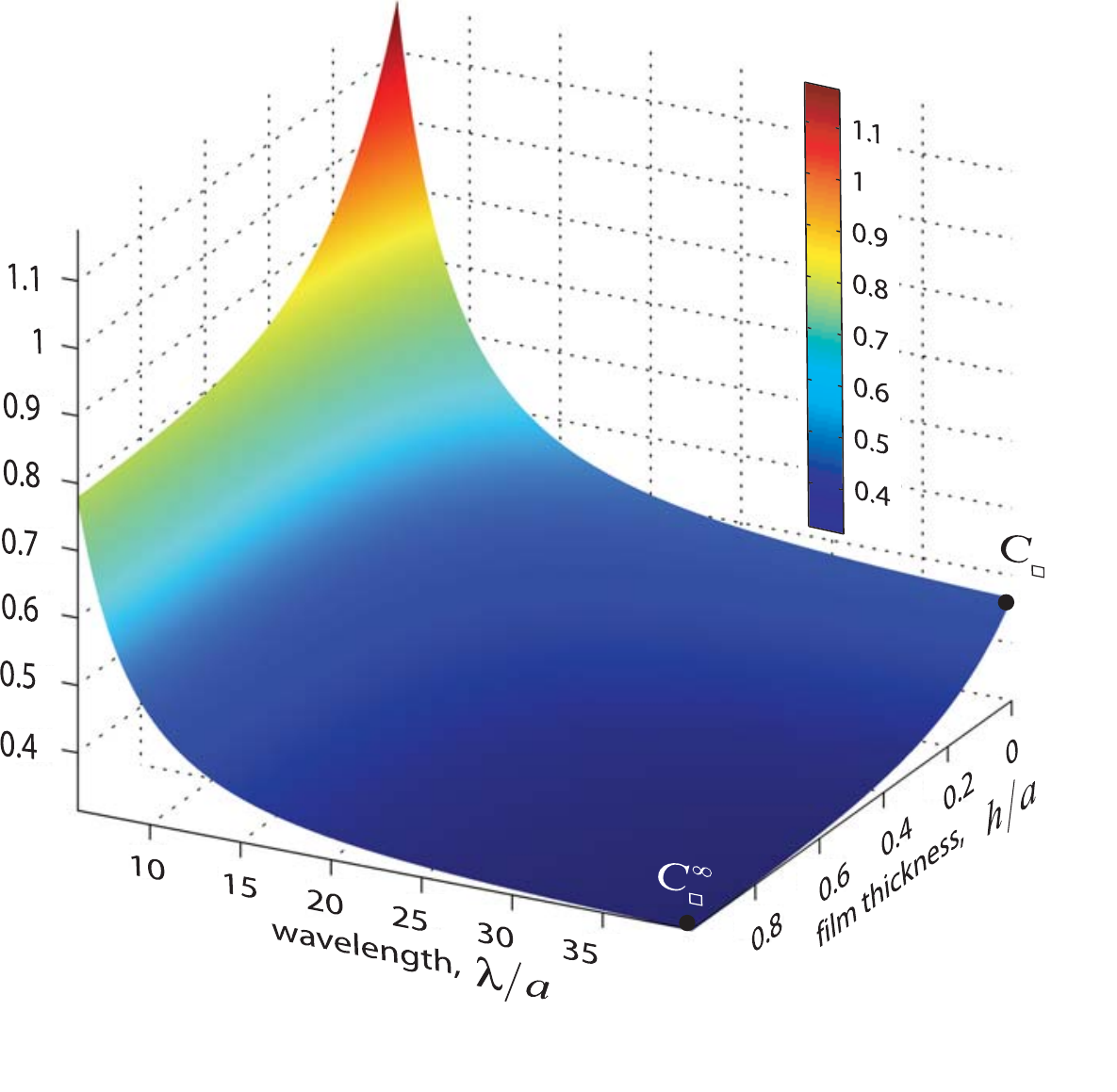}\\
  \caption{(Color online) The normalized transmission through small rectangular hole,
$C(h)+\varepsilon^2C_q(h)$ as a function of the film thickness, $h/a$ and the wavelength, $\lambda/a$, in the units of
the hole half-side.}\label{surface3D}
\end{figure}

The adjusting constants are $C_{2\square}\simeq 0.66$ and $C^\infty_{2\square}\simeq 0.43$. This dependency provides
the transmission with an error not exceeding a few percents in the region $a/\lambda<1/6$ (i.e. $\varepsilon<\pi/3$),
and for an arbitrary film thickness $h$. In the limit of the zero thickness screen, Eq.~\eqref{ms1} has a form similar
to Bouwkamp's expansion (up to the sixth-order term).\cite{Bouwkamp54} However, $C_2$ is not a constant defining the
sixth-order term of the authentic expansion over $\varepsilon$. The latter term in the real expansion only refines the
value of the transmittance in the small hole limit, meanwhile $C_2$ results from the fitting of the spectrum captured
in a wider region.

\section{\label{sec:multipole}The hole as a multipole}

It is well known that the EM field of any source can be considered as resulting from a superposition of multipoles (see
e.g. Ref.~\onlinecite{Book_VanBladel} and references therein). Recently this viewpoint has been used to study the
transmission properties of collections of holes.\cite{GAbajoPRE05} In this section we establish the connection between
the mode-matching formalism and the multipole expansion. For this purpose let us focus our attention on the far-field.
The in-plane components of the electric field are given by Eqs.~\eqref{tb1}, and its $z$-component can be derived from
Maxwell's equations by a straightforward differentiation. The expression for the far-field can be derived with the help
of the scalar free-space Green function associated to the Helmholtz equation in three-dimensions.\cite{JorgePRL2004}
Additionally, in the region of transmission, the expression for the $\mathbf{E}$ far field, can be computed bu using
Green's function identities.\cite{Book_Jackson} The expression for  $\mathbf{E}$ in terms of the integration of the
field at the face of the hole $\mathbf{E}_t(\boldrm{r}_t)$ over the hole area is
\begin{eqnarray}\label{m1}
\mathbf{E}_{III}^{far}(\mathbf{r}) = \frac{ig}{2\pi}\frac{e^{igr }}{r} \mathbf{u}\times \int\limits_S dS'\, \mathbf{n}
\times\mathbf{E}'_t(\boldrm{r}'_t)e^{- ig\mathbf{u}\mathbf{r}'},
\end{eqnarray}
where $\mathbf{u}$ is the unit vector pointing into the observation point, $\mathbf{u}=\mathbf{r}/r$, and $\mathbf{n}$
is the external normal. Eq.~\eqref{m1} has the form of retarding potentials\cite{Book_Jackson} resulting from the
charges induced by the electric field on the face of the aperture. The field an the face of the hole $z=h$, is
expressed through the waveguide mode amplitudes
\begin{eqnarray}\label{m2}
\mathbf{E}'_t(\boldrm{r}_t) = -\sum\limits_{\alpha}E'_\alpha \langle \mathbf{r}_t|\alpha\rangle.
\end{eqnarray}
In the reflection region, I, the scattered far-field has a form similar to Eq.~\eqref{m1}, but in terms of the field on
the interface $z=0$, i.e., $-E'_\alpha$ is replaed by $E_\alpha$ in Eq.~\eqref{m2}.

If we expand the exponent in the integral of Eq.~\eqref{m1}, $e^{- ig\mathbf{u}\mathbf{r}'}= 1-
ig\mathbf{u}\mathbf{r}'+...$, the far-field can then be written in the form of \emph{effective
multipoles}\cite{Book_VanBladel}
\begin{eqnarray}\label{m3}
\mathbf{E}^{far}(\mathbf{r}) = g^2 \frac{e^{igr}}{r}(\mathbf{u}\times\mathbf{p} + \mathbf{m}
+\nonumber\\
\frac{ig}{2}\mathbf{u}\cdot\hat{\mathbf{Q}}_m+...)\times\mathbf{u},
\end{eqnarray}
where $\mathbf{p}$ and $\mathbf{m}$ are the effective electric and magnetic dipole moments respectively,
$\hat{\mathbf{Q}}_m$ is the effective magnetic quadrupole tensor, etc. Comparing Eq.~\eqref{m3} with Eqs.~\eqref{m1}
and \eqref{m2}, we conclude that \emph{each waveguide mode can be interpreted as a superposition of effective
multipoles}; the far field then results from the contribution of the multipoles of all the waveguide modes. For
example, the effective magnetic and electric dipole moments of the waveguide mode $|\alpha\rangle$ read
\begin{eqnarray}\label{m4}
&&\mathbf{m}_{\alpha} = \frac{1}{2\pi ig}\int\limits_S
dS'\,\,\mathbf{n}\times\langle \mathbf{r}'_t|\alpha\rangle, \nonumber\\
&&\mathbf{p}_{\alpha} = \frac{1}{2\pi }\int\limits_S dS'\,\,\mathbf{r}_t'\times[\mathbf{n}\times\langle
\mathbf{r}'_t|\alpha\rangle].
\end{eqnarray}
Denoting the dipole moments of the reflection region by $\mathbf{p}$ and $\mathbf{m}$, and those of the transmission
region by $\mathbf{p}'$ and $\mathbf{m}'$, we write them as
\begin{eqnarray}\label{m5}
&&\mathbf{m} = \sum_\alpha E_\alpha\mathbf{m}_{\alpha}, \q
\mathbf{p} = \sum_\alpha E_\alpha\mathbf{p}_{\alpha}, \nonumber\\
&&\mathbf{m}' = -\sum_\alpha E'_\alpha\mathbf{m}_{\alpha}, \q \mathbf{p}' = -\sum_\alpha E'_\alpha\mathbf{p}_{\alpha}.
\end{eqnarray}
Thus the hole of a finite thickness can be considered as a coupler between induced multipole moments at both faces. For
example the systems of Eqs. \eqref{m5} and \eqref{tb5} describe the coupling between dipole moments that is enough in
the low order approximation, when the hole is small. We would like to stress that the effective dipole moments are
coupled through all the waveguide modes presented inside the hole. For thick films, however, the contribution of the
fundamental waveguide mode dominates over the other modes.
\begin{figure}[h!]
  \includegraphics[width=7cm]{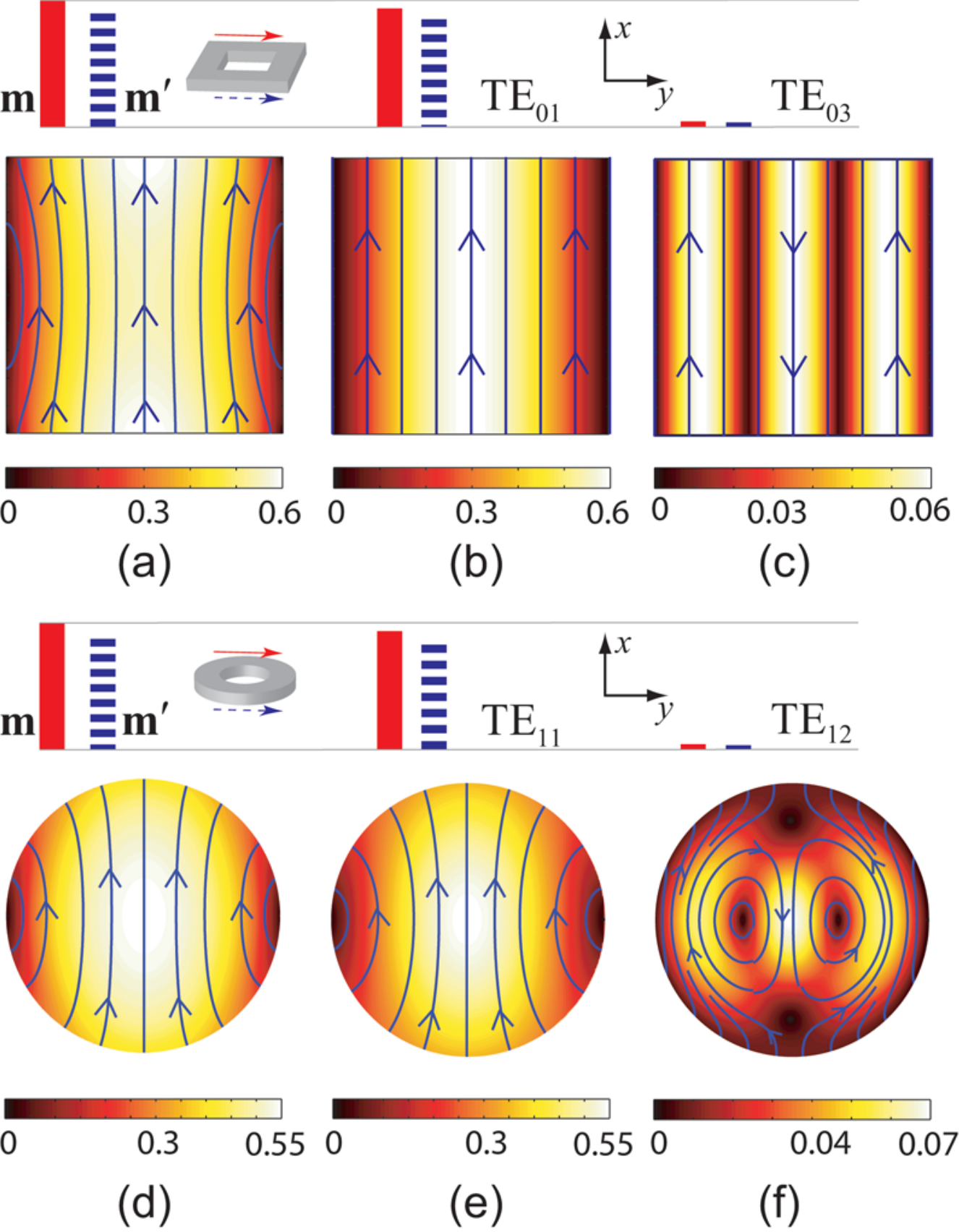}\\
  \caption{(Color online) The magnetic dipole moments and the electric field amplitude
spatial distribution (at $z = h/2$) for the square and circular holes. The amplitudes of the dipole moments are shown
by the bars. The radius of the circular hole is $a$, and the side of the square hole is $2a$. The considered PEC slab
has thickness $h = 0.1a$. The  distribution of the total electric field modulus together with the electric field lines
are shown in (a) and (d). The contribution into magnetic moments from the two lowest waveguide modes and the fields of
these waveguide modes are also shown for both hole shapes: $TE_{01}$ and $TE_{03}$ of the square are represented in (b)
and (c); $TE_{11}$ and $TE_{13}$ of the circle are shown in (e) and (f).}\label{fields}
\end{figure}

Only waveguide modes with zero effective electric dipole moment couple to normal-incident light and generate the
far-field. For the circular hole calculations using Eq.~\eqref{m4} show that only ``horizontal'' $TE_{1n}$ waveguide
modes with integer $n$ contribute to the magnetic dipole moment. For the rectangular hole the contributing waveguide
modes are $TE_{0m}$ and $TE_{m0}$ with odd $m$. When the incident electric field is directed along $Ox$ (magnetic field
along $Oy$), the magnetic dipole moment induced on the incoming face of the hole with the area $S$ is\footnote{Recall
that with our normalization the flux of the incident wave through the hole is unity so that the normalized-to-area
transmittance is expressed trough $\mathbf{m}$ as $T=(4\pi g^4/3)|\mathbf{m}|^2$.}
\begin{eqnarray}\label{m6}
\mathbf{m} =  \mu S\mathbf{e}_y, \q \mathrm{where} \q \mu = \sum_\alpha\tilde{E}_{\alpha}C_{\alpha}.
\end{eqnarray}
We obtain that for the circular and rectangular holes, the constants $C_\alpha$ have the values
\begin{eqnarray}\label{m7}
C^{circ}_{n} = \frac{1}{i\sqrt{2\pi^3(u_{n}^2-1)}}, \q C^{rect}_{m} = \frac{\sqrt{2\tau}}{im\pi^2},
\end{eqnarray}
where $n$ is integer and $m$ is odd. For the dipole moment $\mathbf{m}'$ induced on the outgoing face of the film
$\tilde{E}_\alpha$ must be changed to $-\tilde{E}'_\alpha$. We have checked that the value for $\mu$ of the circular
hole in the limit of zero-thickness screen (that we obtain with the help of 50 modes), $\mu_\circ=0.07624$, reproduces
the Bethe's value $4/(3\pi^{5/2})\simeq 0.07622$. For the square hole we have found $\mu_\square=0.08253$.

Fig.~\ref{fields} renders the effective dipole moments and the exact fields of the small circular and rectangular hole
in the PEC slab of the finite width. The contribution from the fundamental waveguide mode is dominant, so that the
field distribution inside the hole is very similar to that of the fundamental waveguide mode. Interestingly, the vector
lines of the full field are similar for square and circular holes.

It must be noted that in our method the effective dipoles must be computed after solving the systems of
Eqs.~\eqref{tb5},\eqref{m5}. We have been unable to derive the equations governing the effective dipoles directly,
without previous calculation of the amplitudes $E_\alpha$, $E'_\alpha$.

\section{\label{concl}Conclusions}

To conclude, this paper has explored the EM transmission through both small and medium-size isolated holes in a perfect
electric conductor screen of arbitrary thickness. We have used the modal expansion and have shown that this technique
is applicable even in the limit of zero film thickness. This latter limit has been used to check the correctness of the
theory through the comparison with Bethe's result. We have phenomenologically fitted the transmittance in a wide region
of parameters by simple analytical functions.

We have connected the formalisms based on either modal or multipole expansions, and showed that the induced dipole
moments are coupled via all the waveguide modes inside the hole. Our results indicate that the fundamental
(lowest-order) waveguide mode possesses the largest dipole moment. It is also responsible for the attenuation of the
transmission with the increase of the film thickness, $T\sim e^{-2|q_{z0}|h}$.

\section{Acknowledgements}

The authors acknowledge support from the Spanish MECD under contract MAT2005-06608-C02 and Consolider Project
``Nanolight''.

\appendix

\section{\label{GT}Computation of the Green's tensor}

Here we give explicit expressions for the tensor $G_{\alpha\beta}$ in the small hole limit. When the parameter
$\varepsilon$ is small, the integral in Eq.~\eqref{tb8} can considerably be simplified. We illustrate these
simplifications on the example of the diagonal element for the fundamental mode of the square hole with side $2a$. This
diagonal element has the following form in polar coordinates ($q_x = q\cos\theta$, $q_y = q\sin\theta$, $\mathbf{q} =
\mathbf{k}/g$)
\begin{eqnarray}\label{AI1}
&&G_{TE_{01}TE_{01}} = 8i\int\limits_{0}^{\pi/2} d\theta\int\limits_{0}^\infty dq\frac{1 - q^2 \sin^2\theta}
{q\sqrt{1-q^2}}\times\nonumber\\
&&\frac{[1-\cos(2\varepsilon q\cos\theta)][1+\cos(2\varepsilon
q\sin\theta)]}{\left(4\varepsilon^2q^2\sin^2\theta-\pi^2\right)^2\cos^2\theta}.
\end{eqnarray}
We have taken into account the parity of the integrant both in $q_x$ and $q_y$, and, therefore, integrated over the
first quadrant of the $q_x$-$q_y$ plane only.

We see that the integrant in Eq.~\eqref{AI1} is either purely real or purely imaginary depending only upon the square
root $\sqrt{1-q^2}$. Therefore, the integral over $q$ can be separated into two integrals, one from $0$ to $1$
(yielding the imaginary part of $G_{TE_{01}TE_{01}}$) and the other from $1$ to $\infty$ (yielding the real part of
$G_{TE_{01}TE_{01}}$). Then we expand the integrant for the imaginary part into a series over the parameter
$\varepsilon$, retain the leading-terms only and take the integral analytically. The result is
$\mathrm{Im}(G_{TE_{01}TE_{01}})\simeq32\varepsilon^2/(3\pi^3)$. To treat the integral for
$\mathrm{Re}(G_{TE_{01}TE_{01}})$, we make the change of the variable $\xi=2q\varepsilon$.  Taking into account that
the region $\xi\lesssim2\varepsilon$ weakly contribute to the integral, we simplify the square root as
$\sqrt{1-q^2}\simeq iq$. Then we replace the lower limit $\xi=2\varepsilon$ for $\xi=0$ and approximate $1 - \xi^2
\sin^2\theta/\varepsilon^2$ by $- \xi^2 \sin^2\theta/\varepsilon^2$ in the denominator, so that the real part of
$G_{TE_{01}TE_{01}}$ becomes
\begin{eqnarray}\label{AI2}
&&\mathrm{Re}(G_{TE_{01}TE_{01}}) \simeq -\frac{4}{\varepsilon}\int\limits_{0}^{\pi/2}
d\theta\tan^2\theta\int\limits_{0}^\infty d\xi\times\nonumber\\
&&\frac{[1-\cos(\xi\cos\theta)][1+\cos(\xi\sin\theta)]}{\left(\xi^2\sin^2\theta-\pi^2\right)^2}.
\end{eqnarray}
In order to perform the integration over $\xi$ analytically, we extend the integrant into the complex plane, changing
the trigonometric functions of $\xi$ to the exponential functions. Finally, Eq.~\eqref{AI2} is derived applying the
Residue theorem, by taking into account the presence of the poles on the real axis $\xi$. The integral over $\theta$ is
taken numerically.

The other non-vanishing tensor elements both for circular and rectangular holes are simplified analogously. Note,
however, that for a circular hole it is more convenient to perform the analytical integration over $\theta$ in the real
part of $G_{\alpha\beta}$, using the identities for the Bessel functions. The integral over $\xi$ can then be performed
numerically.

In the two following subsections we give simplified Green's tensor elements for the rectangular and circular holes.

\subsection{\label{AIch}Circular hole}

Consider a circular hole of radius $a$. The imaginary part of the Green tensor is approximated as
\begin{eqnarray}\label{AIc1}
\mathrm{Im}(G_{TE_{1m}TE_{1m'}}) = \frac{2\varepsilon^2 }{3\sqrt{(u_{m}^2-1)(u_{m'}^2-1)}},
\end{eqnarray}
where $u_{m}$ are the solutions of the equation $J'_1(u_{m}) = 0$.  After performing the integration over $\theta$, the
real part reads
\begin{eqnarray}\label{AIc2}
&&\mathrm{Re}(G_{TE_{1m}TE_{1m'}}) = \frac{2}{\varepsilon\sqrt{(u_{m}^2-1)(u_{m'}^2-1)}}\times\nonumber\\
&&\int\limits_0^\infty d\xi \frac{\left[\xi J_0(\xi)-J_1(\xi)\right]^2}{\left[1-\left(\frac{\xi}{u_{m}}\right)^2\right]
\left[1-\left(\frac{\xi}{u_{m'}}\right)^2\right]}.
\end{eqnarray}
The right-hand side term of Eq.~\eqref{tb5} is
\begin{equation}\label{AIc3}
\begin{split}
I_{TE_{1m}} = 2i\sqrt{\frac{2}{u^2_{m}-1}}.
\end{split}
\end{equation}

\subsection{\label{AIrh}Rectangular hole}

For a rectangular hole with the sides $2a_x$ and $2a_y$ the imaginary part of $G_{\alpha\beta}$ simplifies to
\begin{eqnarray}\label{AIr1}
\mathrm{Im}(G_{TE_{nm}TE_{n'm'}}) = \delta_{n,0}\delta_{n',0}\frac{32\varepsilon_x\varepsilon_y} {3mm'\pi^3},
\end{eqnarray}
where $\varepsilon_{x,y} = a_{x,y}g$. The real part takes the following form
\begin{eqnarray}\label{AIr2}
\mathrm{Re}(G_{TE_{nm}TE_{n'm'}}) = \frac{1}{\varepsilon_x}\int\limits_{0}^{\pi/2} d\theta\int\limits_{0}^\infty
d\xi\Phi_{nm;n'm'}(\xi,\theta),
\end{eqnarray}
where
\begin{widetext}
\begin{eqnarray}\label{AIr3}
\Phi_{nm;n'm'}(\xi,\theta) = -\frac{2\sigma_n\sigma_{n'}\sqrt{(n^2+m^2\tau^2)(n'^2+m'^2\tau^2)}
\xi^4[1-\cos(\xi\cos\theta)][1+\cos(\xi\tau^{-1}\sin\theta)]}
  {\sin^2\theta\cos^2\theta\left[\xi^2-\left(\frac{n\pi}{\cos\theta}\right)^2\right]
\left[\xi^2-\left(\frac{n'\pi}{\cos\theta}\right)^2\right]
\left[\xi^2-\left(\frac{m\pi\tau}{\sin\theta}\right)^2\right]
\left[\xi^2-\left(\frac{m'\pi\tau}{\sin\theta}\right)^2\right]}.
\end{eqnarray}
\end{widetext}
If $n=0$ then $\sigma_n = \sqrt{2}$ and $\sigma_n = 2$ otherwise. To simplify the double integral \eqref{AIr2} we
transform the cosines in the nominator of $\Phi$ into exponential functions. The integration over $\xi$ is performed
analytically, extending the integrand into the upper complex half-plane, and using the Residue theorem. The
illumination term of Eq.~\eqref{tb5} is
\begin{equation}\label{AIr4}
\begin{split}
I_{TE_{nm}} = \delta_{n,0}\dfrac{4i\sqrt{2}}{m\pi}.
\end{split}
\end{equation}


\begin{thebibliography}{99}
\bibitem{Bethe} H. A. Bethe,  Phys. Rev. \textbf{66}, 163 (1944).
\bibitem{Book_VanBladel} J.~G.~Van Bladel, {\it Electromagnetic Fields} (Wiley-IEEE, New Jersey, 2007).
\bibitem{Book_Jackson} J.~D.~Jackson, {\it Classical Electrodynamics} (Academic,  New York, 1998).
\bibitem{Ebbesen98_Nature} T.~W.~Ebbesen,  H.~J.~Lezec,
H.~F.~Ghaemi, T.~Tio, and P.~A.~Wolff, Nature \textbf{391}, 667 (1998).
\bibitem{Roberts87} A. Roberts, J. Opt. Soc. Am. A \textbf{4}, 1970 (1987).
\bibitem{GAbajoOptExpr02} F.~J.~Garc\'{i}a de Abajo, Opt. Express \textbf{10}, 1475 (2002).
\bibitem{WannemacherOptCom01} R. Wannemacher, Opt. Commun. \textbf{195}, 107 (2001).
\bibitem{PopovApplOpt05} E.~Popov,  N.~Bonod, M.~Neviere, H.~Rigneault,
P.-F.~Lenne,  and P. Chaumet, Appl. Opt. \textbf{44}, 2332 (2005).
\bibitem{SchatzOptExpr05} S-H.~Chang, S.~K.~Gray,  and G.~C.~Schatz, Opt. Express \textbf{13}, 3150 (2005).
\bibitem{ShalaevSarych05} C.-W. Chang, A. K. Sarychev, and V. M. Shalaev, Laser Phys. Lett. \textbf{2}, 351 (2005).
\bibitem{Ebbesen_singHole_OptCom04} A.~Degiron, H.~J.~Lezec, N.~Yamamoto, and T.~W.~Ebbesen,
Opt. Commun. B \textbf{239}, 61 (2004).
\bibitem{hole_shape_PRL04} K.~J.~K.~Koerkamp, S.~Enoch, F.~B.~Segerink, N.~F.~Hulst, and L.~Kuipers,
Phys. Rev. Lett. \textbf{92}, 183901 (2004)
\bibitem{hole_shape_PRL07} J.~W.~Lee, M.~A.~Seo, D.~H.~Kang, K.~S.~Khim, S.~C.~Jeoung,
 and D. S. Kim, Phys. Rev. Lett. \textbf{99}, 137401 (2007)
\bibitem{rect_hole_PRL05} F. J. Garc\'{i}a-Vidal, E. Moreno, J. A. Porto, and L. Mart\'{i}n-Moreno,
Phys. Rev. Lett. \textbf{95}, 103901 (2005).
\bibitem{resonant_hole_PRB06} F. J. Garc\'{i}a-Vidal, L. Mart\'{i}n-Moreno, E. Moreno, L.~K.~S.~Kumar, and R.~Gordon,
Phys. Rev. B \textbf{74}, 153411 (2006).
\bibitem{JorgePRL2004} J. Bravo-Abad, F. J. Garc\'{i}a-Vidal, and L. Mart\'{i}n-Moreno,
Phys. Rev. Lett. \textbf{93}, 227401 (2004).
\bibitem{Polariz_exper_IEEE} S. B. Cohn, Proc. of IEEE \textbf{39}, 1416 (1951).
\bibitem{Polariz_teor_IEEE} F. de Meulenaere and J. V. Bladel, IEEE Trans. on Ant. and Prop.
\textbf{25}, 198 (1977).
\bibitem{Bouwkamp54} C. J. Bouwkamp, Rep. Prog. Phys. \textbf{17}, 35 (1954).
\bibitem{GAbajoPRE05} F.~J.~Garc\'{i}a de Abajo, R.~G\'{o}mez-Medina, and J.~J.~S\'{a}enz,
 Phys. Rev. E \textbf{72}, 016608 (2005).





\end{thebibliography}
\end{document}